# A PROPOSED FRAMEWORK FOR THE COMPREHENSIVE SCALABILITY ASSESSMENT OF ICTD PROJECTS

Gugulethu Baduza, Rhodes University, g.baduza@ru.ac.za

Caroline Khene, De Montfort University, caroline.khene@dmu.ac.uk

**Abstract:** The scalability of ICTD projects is an imperative topic that has been neglected in the field. Little has been written or investigated about the assessment of the scalability of ICTD projects due to factors, such as the lack of proven business models for success, the high failure rate of projects, undefined aspects of assessment, and the small number of projects that have scaled. Therefore, there are various factors that should be taken into consideration to alleviate the challenges experienced in the process of scaling up. This research study is guided by an investigation into how can the scalability of an ICTD project be assessed using a comprehensive evaluation approach that considers the impact and potential sustainability of the project. This research study proposes a Comprehensive Scalability Assessment Framework (CSAF), using systems theory and amplification theory to guide the theoretical analysis and empirical investigation. A theorizing approach is used to develop the framework, which is structured around three components: *assessment guidelines and proceeding domains of evaluation*; *four scalability themes (stakeholder composition, models feasibility, resources sustainability and resilience)* and *judge scalability*.

## 1. INTRODUCTION

Social pilot programmes that are targeted at societal development, according to Toyama (2015), usually have no impact because of their nature and are scaled when they should not have scaled, usually for three reasons. These are bad program design which, firstly, includes 'a problem with the theory of the intervention'; secondly, with 'how the theory was implemented'; and the third reason relates to faulty implementation which would result in the project failing at scale (Toyama, 2015). All these reasons provide the greatest motivation as to why a project should be evaluated before the project proceeds to scale. In the Batchelor and Norrish (2007) framework for the assessment of an ICT pilot project, the outcomes and data from the project purpose assessment provide a base on which questions can be asked as to how scale will happen under similar conditions in the next environment.

Scaling up has multiple definitions, but it is generally agreed that scaling up means the expansion, adaptation, replication and sustaining of desired policy, programme and practice changes (World Bank, 2012; Walsham, Robey and Sahay, 2007; Gerhan and Mutula, 2007; Batchelor and Norrish, 2007). Implied in the definitions of scaling up is the assumption that we scale up in order to expand valued outcomes, such as poverty reduction, or meeting the goals of the country and community, and any World Bank strategies (World Bank, 2012). Up-scaling also implies increasing benefits to other communities, where they can access the services provided by the up-scaled project. A participative ICTD approach makes a difference when scaling-up, this is the case when it involves people based on the identification of their needs and a collaborative assessment to monitor the outcome of the project (Gerster and Zimmermann, 2005).

## 2. COMPREHENSIVE SCALABILITY ASSESSMENT OF ICTD PROJECTS

The process of conducting a scalability assessment is a process that determines whether a project should be scaled-up or not. Theorizing, according to Weick (1989), is a process that assists in developing a disciplined approach to structure the imagination of the people or person developing the theory, as it allows for more options to strengthen the process of theorizing. The key elements of this process to strengthen theorizing are entrenched in being able to identify and understand the relationships, interdependencies and the connections in the subject being investigated (Weick, 1989). The process involved three stages which are problem formulation, thought trials and selection criteria which informed the process.

The results of the theorizing process, where guided by a critieria that is in table 1.1. The development of the '*Comprehensive Scalability Assessment Framework*' is based on the review of 9 frameworks that are aimed at conducting a scalability assessment in fields which contribute to ICTD. Critical themes such as stakeholder composition, resource sustainability, resilience and model feasibility have been identified as key points in the assessment process, and have contributed to the development of this proposed framework. The concept of comprehensive evaluation becomes almost non-existent in reference to ICTD pilot projects. The focus of such ICTD pilot projects tends to be on the broad outcomes of the initiative, in relation to the tangible and quantifiable indicators that are pleasing to the external stakeholders. The need for a comprehensive evaluation of ICTD pilot projects has increased so as to ascertain the outcomes of the project and their scalability or replicability to the existing project or to a new contextual environment.

| *Criteria* | *Description* |
|---|---|
| *1. Focus is on conducting a scalability assessment* | The framework should focus on assessing if a programme or project is scalable and should indicate what methods were used to determine if it is ready to scale. |
| *2. Indicates what is assessed in the project* | The aspects that are assessed are mentioned and the methods to access them are available. This will assist in identifying possible themes and how they are assessed. |
| *3. Components can possibly be adapted and applied in ICTD and local context* | Although grouped into categories, the approaches or frameworks have relevance to the ICTD field and can assist in its being applicable to the context and relatable to ICTD. |
| *4. Approach or framework meets criteria in Table 1.1* | The approaches or frameworks can be analysed based on the template in Table 6.4 that will be used to analyse the selected approaches or frameworks. |
| *5. Evidence of mixture of theoretical propositions and case study application* | As theory will be proposed in the approach or framework, it will inform the themes considered, but the practical application in a case study will provide insights into lessons learned, tools, responses, etc, that will contribute to the proposed framework. |

**Table 1.1 Criteria used for selecting approaches or frameworks to be analysed**

Crucial to the comprehensive evaluation of ICTD pilot projects is the basis of understanding the need for such a program through the form of a needs assessment, in order to solidify the motivation for having such a project in relation to the prototype and to the social agenda of the project. The comprehensive evaluation needs to encompass a programme theory assessment, process assessment and outcome assessment in order to contribute to a clearer scalability assessment that encompasses these elements. The focus of this paper will be introduce a proposed approach to conduct a scalability assessment based on a comprehensive evaluation of an ICTD pilot project. Enabling this focus will involve unpacking the structure and the composition of the framework

A scalability assessment aims to review various areas and elements in a programme to determine if the programme is ready to scale (Cooley and Kohl, 2012). The desired outcome of the process is for the review of the various frameworks to work towards developing aspects that could contribute to the framework. As indicated by Weicks (1999), it is a trial-and-error process that is necessary, in the theorising process, to develop a commonality amongst the concepts of assessing for scalability. The review of middle range theories supports the process of thought trials that assist in the development of a scalability assessment framework. Table 1.2 is a review of the frameworks that were selected and reviewed in the development of the Comprehensive Scalability Assessment Framework.

| *Category* | *Approach or Framework* | *Description* |
|---|---|---|
| ***Social Programmes*** | Framework 1: Scaling Up: From Vision to Large-scale Change, A Management Framework for Practitioners (Cooley and Kohl, 2012). | This scaling management framework is based on a three-step ten-task approach, which aims to facilitate effective up-scaling based on theory and practice from the field. |
| | Framework 2: Institutional Approaches and Organizational Paths To Scaling Up (Hartmann and Linn, 2008). | This framework is aimed at enabling the key dynamics that allow the scaling-up process through exploring the possible approaches to scale, drivers of replication, space to be created for scaling, and the role of careful planning and implementation. |
| | Framework 3: Lessons from Practice: Assessing Scalability (Holcombe, 2012). | The framework aims to establish proof of innovation impact and scalability, and sufficient information to assist in the process of determining if the project should be scaled up or not. |
| | Framework 4: Scaling-Up the Impact of Good Practices in Rural Development (Hancock, 2003). | The framework is centred around identifying and classifying the information based on what has worked, and what has not worked, according to the evidence that is available to the reviewers. |
| ***IT/IS Programmes*** | Framework 5: Information Infrastructures (Rolland and Monteiro, 2002). | The aim of the theoretical framework is to guide the development of Information Infrastructures which are interconnected and integrated and which use comprehensive modules of Information systems to extend operations across many contexts. |
| | Framework 6: A framework for the Evaluation and Justification of IT/IS (Gunasekaran, Ngai and McGaughey, 2006). | The aim of this framework is to provide an IT/IS justification process that would assist e-businesses to make well considered decisions which take into account the goals, |

| | | costs, and new environmental complexities. |
|---|---|---|
| | Framework 7: Project Success, A Multidimensional Strategic Concept (Shenhar, Dvir, Levy and Maltz, 2001). | The framework is based on defining and assessing project success in order to align the short-term and long-term goals of the organisation in order to enable the strategic development of technology projects. |
| *ICTD Programmes* | Framework 8: Framework for the assessment of ICT pilot projects (Batchelor and Norrish, 2007). | The aim of the framework is to conduct M&E for pilot projects, to assess developmental purpose and proof of concept, and to provide recommendations that enable scale based on the assessment. |
| | Framework 9: Rural ICT-Comprehensive Evaluation Framework (RICT-CEF) (Pade-Khene and Sewry, 2011). | The aim of the framework is to conduct a comprehensive evaluation of rural ICT projects, through an iterative process and through various domains, in order to evaluate the project. |

**Table 1.2 Approaches and Frameworks Selected**

## 2.1. THE STRUCTURE OF THE COMPREHENSIVE SCALABILITY ASSESSMENT FRAMEWORK

As shown in Figure 1, the framework is designed in a cyclic manner that has various components attached to it. Keeping in line with systems theory, the framework can be viewed as a system with subsystems. The framework is divided into three main areas which are the assessment guidelines and preceding domains of evaluation, with the four themes and judge scalability as the final stage. The assessment guidelines are placed on the outside of the inner circle, because they are meant to guide every action that is taken in the various subsystems. This is the same for the preceding domains of evaluation, which is a process of providing proof that the project has been evaluated and providing insight into what the outcome of each assessment was. The second part of the framework is the composition of the four stages that work towards the judge scalability component. These themes lead to each other and cannot be executed in isolation of each other, as they work in a cyclic manner to contribute to judging scalability, as shown in the diagram. These themes are the stakeholder composition, model feasibility, resource sustainability, and resilience. The dashed lines mean that fluidity between the stages is possible, and flexibility is encouraged, especially when the process of judging for scalability has started.

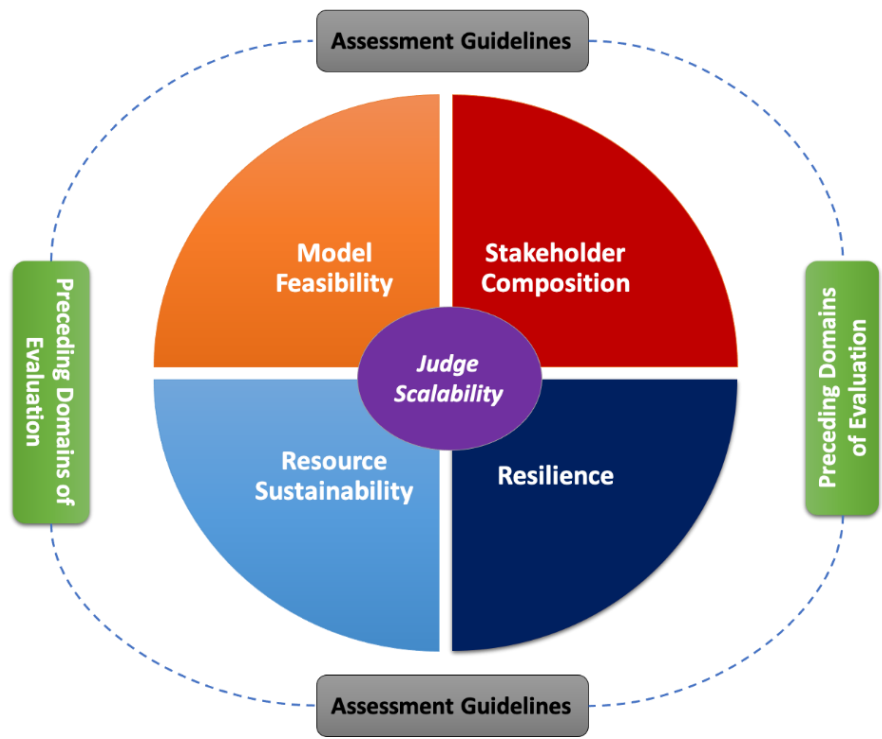

**Figure 1: Comprehensive Scalability Assessment Framework**

These themes are the stakeholder composition, model feasibility, resource sustainability, and resilience. The dashed lines mean that fluidity between the stages is possible, and flexibility is encouraged, especially when the process of judging for scalability has started. All these factors are driven by the concepts of systems theory and amplification theory, which influence the design of the framework. This includes incorporating factors that present a real-world view in a holistic manner and take into account the environment in which the system operates. Additionally, contextual factors that influence the viability of the system are considered, in line with amplification theory. Systems theory allows us to understand how a system functions to achieve a goal in a complex situation with diverse stakeholder involvement and various objectives and goals. Amplification theory supports a deeper investigation to understand what existing forces in the system enable the various subsystems to function so that the system operates on a daily basis.

## 2.2. CONDUCTING A COMPREHENSIVE SCALABILITY ASSESSMENT OF ICTD PROJECTS

The process of conducting a scalability assessment is a *pragmatic judgement* process that aims to review the suitability of a model and the factors in the larger context which affect the success of the implemented project (Cooley and Kohl, 2012; Keijdener, Overbeek and Espãna, 2017). The implication of this pragmatic process is that it is dependent on comprehensively evaluating the application and impact of the selected model.

### 2.2.1. Assessment guidelines

Assessment guidelines guide and contribute to the overall application of the framework. These guidelines determined key areas that needed to be adhered to prior to the application of the framework. These are the application of a comprehensive evaluation, the stakeholder relationship and ethical considerations.

- *Application of Comprehensive Evaluation as the Base of the Scalability Assessment Framework:* In order to provide credible evidence and to commence a scalability assessment, proof must be given to relevant stakeholders of the viability of the project and the impact it has had. This means that various evaluations of the project need to be conducted on all aspects of the project.

- *Stakeholder Relationships*: Stakeholder management is a concept that has been addressed by many ICTD studies and is usually highlighted as one of the elements that 'make or break' projects, depending on how well or how badly their role in a project is emphasized (Pandey and Gupta, 2017). Stakeholder engagements usually start off with stakeholders being consulted on a decision affirming point and not being actively involved in making the decisions of the project.
- *Ethical Considerations*: It is important that ethical guidelines are presented to all stakeholders to ensure that the actions conducted within the study are not seen as improper, unlawful and unacceptable in the community (Traxler, 2012). Transparency, communication, and feedback should govern the process of engaging stakeholders in the scalability assessment of ICTD projects. It should also be communicated clearly to the various stakeholders what benefit the assessing of scalability provides for them, thereby making it their choice to be part of the assessment.

### 2.2.2. Preceding Domains of Evaluation

There are a number of domains of evaluation that need to be conducted which will then provide comprehensive information on the status of the programme and how, if possible, it can be scaled-up. Therefore, the decision to scale is not to be taken lightly, and cannot be decided based solely on the information conducted from an impact assessment but from other domains which informed the setup of the project during its inception, and should incorporate the progress made throughout the project.

These are some of the evaluation domains which should be considered and be used to inform the development and implementation of the scalability assessment framework:

- *Baseline study*: The baseline study aims to understand and determine the existing status of the community in relation to its socio-economic status and its ability to be part of the new envisioned programme (Pade-Khene and Sewry, 2011; Batchelor and Norrish, 2007). In the process of a scalability assessment, such information would inform the possibility of different future phases or staggered implementation that could be done in the community, based on various factors such as capacity and skills. If no baseline study is available the use of previous studies, household surveys or other suitable socio-economic studies might be used.
- *Needs Assessment*: A needs assessment is referred to as a process of investigation into the various needs and priorities of the intended community, which are demand driven, with the aim of providing suitable interventions or approaches that might be taken to deal with these needs and priorities (Pade-Khene and Sewry, 2011; Rossi, Freeman and Lipsey, 2004; Pandey and Gupta, 2017; Mthoko and Khene, 2018). In a pilot project, the aim is to address a specific problem that exists. This assessment informs the scalability assessment of other aspects that might need to be expanded on in future and what needs and priorities it might try to address in future.
- *Programme Theory Assessment*: The aim of the programme theory assessment is to provide a plan or a blueprint of the planned intervention based on the needs assessment and to evaluate its suitability towards meeting the needs of the intended beneficiaries (Pade-Khene and Sewry, 2011; Rossi, Freeman and Lipsey, 2004; Pandey and Gupta, 2017; Mthoko and Khene, 2018). It is important for the scalability assessment to take into account the planned model for intervention and to review and verify if it will work in its scaled-up version. Moreover, this assessment would need contextual factors to be incorporated into the plan to scale.

- *Process Assessment*: This assessment assesses and evaluates how the planned programme is being implemented, and evaluates the processes and activities to see if they work towards solving the targeted problem (Batchelor and Norrish, 2007; Gigler, 2004; Pade-Khene and Sewry, 2011; Rossi, Freeman and Lipsey, 2004). In the process of planning a scalability assessment, there needs to be a review of the various processes and activities currently in place to make the project function. This would involve assessing the various resources, skills and capacity to conduct the same activities in a different context, or assessing how the activities and processes would be adjusted to ensure that similar results can be achieved in a different context.
- *Outcome and Impact Assessment*: The process entails ascertaining the intended and unintended effects of the implementation of the programme and contributes to the outcomes and impact assessment of a programme (Pade-Khene and Sewry, 2011; Mthoko and Khene, 2018). It is a process that contributes greatly to understanding how the programme could potentially benefit other contexts for them to reap similar rewards in their own context.
- *Efficiency Assessment*: The aim of this assessment is to assess the various ICT intervention costs associated with projects effects or impact (Pade-Khene and Sewry, 2011; Pandey and Gupta, 2017). This was also a strong contributor to the aspect of assessing for scale in terms of reviewing costs that would be incurred internally and externally.

The outcome of the combined domains of evaluation contributes to the process of answering the how, why and when of the aspects relating to the scalability assessment. Moreover, these contribute to the guidelines and preliminary measures that need to be put in place before the scalability assessment framework is implemented.

### 2.2.3. Judge Scalability
#### 2.2.3.1.    Stakeholder Composition

Assessing scalability, in terms of stakeholder composition, is a process that aims to understand which key and contextual stakeholders are needed for the project to work effectively and deliver the desired results in the selected type (vertical or horizontal) of scaling. To assess and review stakeholder approaches by various frameworks, literature has utilised various approaches, namely descriptive, normative and instrumental, with each having its weakness which renders it impractical (Bailur, 2006). The approach that can be utilised is the concept of Actor Analysis, which aims to provide a high-level view of the various stakeholders in the project at the initial exploration phase (Enserink, Hermans, Kwakkel, Thissen, Koppenjan and Bots, 2010).

#### 2.2.3.2.    Resource Sustainability

The process of assessing resource sustainability aims to interrogate the various forms of resources that can be sustained before the process of scaling up. The aim is to assess whether there are sufficient resources and if they are sustainable, to ascertain if the project will continue. However, keeping in mind the rigorous debate of the topic of sustainability, the aim is to assess the suitability of the resource sustainability to the context of the project going forward. As the theme is broad, in order to assess it, there is a need to review the various types of resource sustainability in relation to the project and its context. For the purposes of this research, the various types of sustainability will be termed sub-themes and are assessed in relation to other themes which all contribute to the final judge scalability aspect. Heeks (2005) provides an approach to reviewing various aspects of sustainability within the project by utilising capacity, utility and embedding to assist in reviewing the various sub-themes. *Capacity* focuses on understanding if there are the skills, data, funds or technology available to continue the project, after the pilot phase (Heeks, 2005; Toyama, 2015). *Utility* is usually a key determinant which focuses on the usefulness of the project to the community, based on the reviewed aspect of sustainability (Heeks, 2005; Ali and Bailur, 2007). *Embedding* focuses on understanding if the project has become routine and institutionalised within the community (Heeks, 2005). When taking into account the context, the variations that exist within communities, and the considerations of achieving bricolage to assist sustainability, a fourth element to reviewing sustainability can be added. *Contextual leverage* is added as an element that focuses on reviewing what other aspects of the project and the context can be used to the advantage of sustaining the project (Ali and Bailur, 2007).

### 2.2.3.3.    Resilience

Resilience is a concept that is not too far from the concept of sustainability and bricolage. Resilience is understood to be the ability of a system to adapt and recover from the shock that it experiences from internal and external factors (Walker and Salt, 2012; Heeks and Ospina, 2018). It is linked to key factors which illustrate resilience as the ability to experience stability, agility, flexibility, adaptation and transformability of the system based on the changes that affect the system (Walker and Salt, 2012; Marias, 2015; Heeks and Ospina, 2018; Chen, 2015). The process of assessing resilience in order to determine scalability should be based on understanding the resilience of the community. Walker and Salt (2012) recommend that there are three aspects to understanding resilience, which include describing resilience, assessing resilience and managing the resilience of the system. For purposes of this research, the focus would be on assessing for resilience; however, that cannot be done without understanding the system's composition and describing the resilience of the community. Therefore, for purposes of assessing for scalability in this framework, it is further broken down into three stages: *Resilience Foundation*, *Assessment of Resilience* and the *Effect of Resilience on Scalability*.

### 2.2.3.4.    Model Feasibility

Assessing the verification, feasibility and transferability of the implemented model is a process that relies heavily on the linkage between the programme theory assessment, process assessment and the outcome and impact assessment (Cooley and Kohl, 2012; Holcombe, 2012). The implemented model would need to be broken down into various parts in order to view each element, with its functions and how they are related, to understand how they work together in a holistic manner, to conclude the project. This means each structure of the model needs to be understood in terms of what structures enabled it to work, which stakeholders were involved in the process, what the environment was that led to the success of the project, and so forth.

### 2.2.3.5.    Judge Scalability

The aim of this stage is to provide feedback on the entire process of assessing for scalability and compiling a report that can be used to determine, based on the information provided, if the project can be up-scaled or not, or if certain factors should be implemented differently for the project to be considered scalable. The report should be provided to all stakeholders including the intended beneficiaries. As this is the final stage, the assessment of the scalability of the project would have been a consultative process, incorporating the various views of all crucial stakeholders (Pade-Khene and Sewry, 2011; Batchelor and Norrish, 2007). The assessment of each theme would be available at this stage, and the outcome would provide evidence of the possibility to scale or not. The interpretation of the outcome is the point where there would need to be a negotiation or consensus of, '*is this enough evidence to scale the project or not scale the pilot project?*'

## 3. CONCLUSION

The process of assessing scalability in ICTD projects should be conducted in a comprehensive manner which takes into account the importance of evaluation, making decisions with the input of all stakeholders, considering all contextual factors, and the assessment of the project. The *Comprehensive Scalability Assessment Framework* is developed based on a critical review of the field and its frameworks. This proposed framework is structured in a manner that uses assessment guidelines throughout the assessment process, incorporates the results of the comprehensive evaluation and then sets in motion a scalability assessment plan that contributes to the final scalability judgement. At the centre of the proposed framework are the four themes which make up the scalability assessment, namely *'Stakeholder composition'*, *'Resource sustainability'*, *'Resilience'*, *'Model feasibility'*, and with '*Judge scalability*' as the end process of the framework. A systems analysis diagram is the main output of the judge scalability stage and is interpreted in order to make a decision to scale the project or not. The proposed framework, therefore, aims to provide a comprehensive report that decides on the scalability of the project based on the information and evidence gathered, to actively decide with all stakeholders if the project should be scaled-up or not. In order to have a clear view and understanding of the suitability, practicality and shortcomings of the proposed framework, it is important to apply the framework in a real-life setting and to use the results of the implementation to revise the proposed framework.